# Modular Simulation Framework for Process Variation Analysis of MRAM-based Deep Belief Networks

Paul Wood, Hossein Pourmeidani, and Ronald F. DeMara

Department of Electrical and Computer Engineering

University of Central Florida, Orlando, FL 32816-2362

pwood9@knights.ucf.edu, hossein.pourmeidani@knights.ucf.edu, and ronald.demara@ucf.edu

*Abstract* — Magnetic Random-Access Memory (MRAM) based p-bit neuromorphic computing devices are garnering increasing interest as a means to compactly and efficiently realize machine learning operations in Restricted Boltzmann Machines (RBMs). When embedded within an RBM resistive crossbar array, the p-bit based neuron realizes a tunable sigmoidal activation function. Since the stochasticity of activation is dependent on the energy barrier of the MRAM device, it is essential to assess the impact of process variation on the voltage-dependent behavior of the sigmoid function. Other influential performance factors arise from varying energy barriers on power consumption requiring a simulation environment to facilitate the multi-objective optimization of device and network parameters. Herein, transportable Python scripts are developed to analyze the output variation under changes in device dimensions on the accuracy of machine learning applications. Evaluation with RBM circuits using the MNIST dataset reveal impacts and limits for processing variation of device fabrication in terms of the resulting energy vs. accuracy tradeoffs, and the resulting simulation framework is available via a Creative Commons license.

*Keywords* —Neuromorphic Computing Hardware, Python-Script Driven Simulation, Magnetic Random-Access Memory (MRAM) Device, Stochastic Neuron, Restricted Boltzmann Machine (RBM).

## I. Introduction and Background

A *probabilistic bit (p-bit)* device is a two-terminal voltage-controlled Magnetic Random-Access Memory (MRAM) component realizing a sigmoidal activation function as an artificial neuron suitable for machine learning applications such as Restricted Boltzmann Machines (RBMs) [1]. A p-bit is formed by combining a common source NMOS transistor with an in-plane MTJ (IMTJ) between $V_{DD}$ and the drain of the NMOS transistor followed by a CMOS inverter. This creates an oscillating voltage at the drain of the NMOS transistor that can then be pinned to $V_{DD}$ when $V_{IN}$ is $V_{DD}$ or alternatively to ground when $V_{IN}$ is below the NMOS threshold voltage. The sigmoid function is realized via the stochastic output of the p-bit due to thermal noise induced switching [1].

Deep Belief Networks (DBNs), specifically Restricted Boltzmann Machines (RBMs), can be physically realized by utilizing a resistive weight crossbar array with a p-bit neuron to realize the sigmoidal activation function. To date, a simulation framework called *Probabilistic Inference Network (PIN-sim)* was developed in order to create, train, and test RBM circuit models [1]. The framework is composed of five blocks:

*1) TrainDBN:* a MATLAB training algorithm

*2) MapWeight:* a MATLAB algorithm used to convert neuron weights and biases into resistor values

*3) MapDBN:* a python script that realizes the weights and biases into a circuit model

*4) Neuron:* the SPICE model of the p-bit

*5) TestDBN:* a python script that tests the error and power consumption of the DBN.

To interpret the results of the probabilistic reasoning process as a definitive selection, a conversion circuit is needed to transform the time-varying stochastic analog signal into a digital output such as a Probabilistic Inference Recoder (PIR). While functionally similar to an Analog Digital Converter (ADC), PIR circuits realize significant area and power consumption reduction of 48% and 74% versus ADCs, respectively [2]. To achieve practical fabrication of machine learning hardware, it is of particular interest to analyze how RBM models using PIR circuits behave in the presence of manufacturing variations prevalent at the 45nm process node and below. Herein, a Python script-based simulation framework was developed to extend PIN-sim that invokes SPICE neuron models at different energy barriers and collates the data into a format suitable for the designer to balance tradeoffs involved.

## II. Python-Driven Simulation Framework

The goal with this script is to gather data on how process variation will affect the energy barrier of the MTJ, thus changing the realization of the sigmoid function and potentially adverse effects to energy consumption. The Python script invokes SPICE to gather outputs under consecutive voltage data points applied to the p-bit device as shown in Figure 1. Given a SPICE neuron file with the small magnetic anisotropy field, $H_K$, defined in the parameter file, the script changes $H_K$ values based on the propagated energy barrier value. It then runs the simulation while piping the `bash` output to a text file in case of any SPICE errors. Finally, it extracts the neuron output voltage data points and collates them within the `results` text file. Multiple energy barrier values can be run sequentially if a text file containing a list of energy barriers with each entry is passed as an argument to the script.

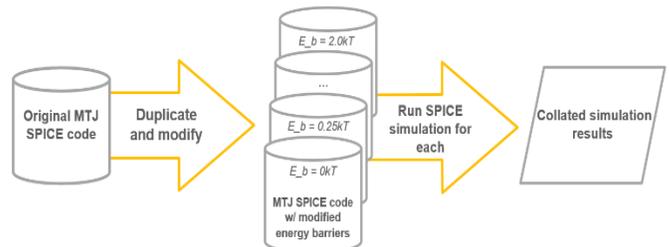

Figure 1: MTJ Energy barrier simulation using Python scripting.

Data points can then be plotted in MATLAB to view the effect that the energy barrier has on the realization of the sigmoid activation function. The formula used by the Python script for calculating the energy barrier, $E_b$, is:

$$E_b = \frac{1}{2} H_K M_s V \quad (1)$$

where $V$ is the volume and $M_S$ is the magnetization saturation of the MTJ. The pseudocode for the script is identified in Algorithm 1.

---

**Algorithm 1:** Effect of Process Variation on the MTJ Sigmoid Activation Function Realization

---

**Input:** energy barriers, neuron source file
**Output:** neuron output voltage

**for each** energy barrier, E:
    **calculate** $H_K = \frac{2 E_b}{M_s V}$
    **search** for "HK= " in SPICE code
    **replace** anisotropy field value with calculated value
    **run** SPICE simulation piping bash output to text file
    **search** for voltage output and **write** to results file

---

Another tool developed was a Python script to aid testing of the DBN's accuracy with the MNIST dataset. The development of these scripts extends PIN-sim and will aid future development and testing of p-bit based, DBN networks with a PIR digitization output stage.

### III. MNIST DATASET EVALUATION

To analyze the performance of the PIR circuit, a Python script was developed to compare the large amounts of data commonly found in machine learning datasets. This accuracy analysis script operated as follows: first, it reads one line of the MNIST dataset file to find the expected output for that testcase and the testcase label. Secondly, it locates each instance of the testcase label in the PIR output file. Third, it reads the neuron data into a list until it encounters the subsequent case. Upon finding the next testcase, all the neurons for the current testcase have been read and now can be processed. The list is sorted by constituent probabilities from high to low. Next, the first two neurons are examined to see if either of them is the neuron indicating the expected output from the MNIST dataset. If any output digit neuron subsequent to the top two neurons have the same probability, then the output of the PIR circuit counts as a `fail` even if the expected output was within the top two confidence selections, whereas the circuit was not able to tell a clear difference between which neuron was correct. If the expected output was a neuron in the top two likelihood categories, then the testcase is regarded as a `pass`.

The process is repeated until either file reaches its end. Upon finishing all testcases, then the total number of testcase passes and failures are tabulated so that the overall error rate is determined. The corresponding pseudocode is listed in Algorithm 2. The script was tested by comparing the error rates generated against previous works [2]. Namely, the script was fed outputs of a PIR circuit with 100 testcases and the MNIST dataset. The results obtained are listed in Table I.

---

**Algorithm 2:** MNIST DBN Performance Analysis

---

**Input:** MNIST dataset, PIR output
**Output:** number of testcases that passed/failed

**for each** testcase
    **for each** neuron
        append neuron data to list
    **sort** list by neuron probability high to low
    **if** the expected output was in the top two neurons
        **AND** its probability doesn't match any neurons
        beyond the top two neurons
    **then**
        testcase **passes**
    **else**
        testcase **fails**

---

Table I: Optimization of precision, area, and accuracy for handwritten digit recognition using MNIST dataset.

| PIR Precision | Energy Consumption | Error Rate |
|---|---|---|
| 3 bits | 90.75 fJ | 24% |
| 4 bits | 124.2 fJ | 17% |
| 5 bits | 176.0 fJ | 18% |

### IV. CONCLUSION

MRAM-based p-bit neuromorphic architectures offer an emerging device approach to realize true intrinsic machine learning within accelerators and IoT edge devices. A simulation framework using Python scripts can thoroughly address two important areas requiring analysis prior to fabrication: effects of the variation on the p-bit sigmoid function and network optimization for accuracy vs area using an appropriate analog signal digitization strategy, such as a PIR circut. Because the p-bit's stochasticity is dependent on oxide barrier thickness, a one script ran a SPICE script multiple times, each changing the magnetic anisotropy field determined by the oxide thickness and thus permuting the energy barrier as a process variation yielding repeatable results to the desired confidence interval. The resulting simulation framework is transportable, adaptable, and available to other machine learning researchers via a Creative Commons license upon request to the authors.

ACKNOWLEDGMENTS

This work was supported in part by the Center for Probabilistic Spin Logic for Low-Energy Boolean and Non-Boolean Computing (CAPSL), one of the Nanoelectronic Computing Research (nCORE) Centers as task 2759.006, a Semiconductor Research Corporation (SRC) program sponsored by the NSF through CCF-1739635.